# Detailed Balance and Spin Content of Λ using Statistical Model

M. Batra, A. Upadhyay
School of Physics and Material Science
Thapar University, Patiala
alka@thapar.edu, mbatra310@gmail.com

**Abstract:**
In this study, we assume the hadrons as an ensemble of quark-gluon Fock states and apply statistical effects to explain the spin distribution of quarks for lambda hyperon. We extend the principle of detailed balance given by Zhang et al. [9] to calculate the probability of every Fock state in lambda. Using these probabilities, we have calculated quark distribution of spin, hyperon weak decay constant, SU(3) reduced matrix elements and their ratio and magnetic moment etc. using the statistical model proposed in ref.[17]. The extension of statistical model comes here with inclusion of more sub processes like $s \Leftrightarrow sg, g \Leftrightarrow s\bar{s}$ etc. Finally the matching of our results with other models like chiral quark solitan model has also been analyzed.

**Introduction:**
The structure and spin of proton has always been of keen interest for experimentalists as well as theoreticians during last few years. Flavor asymmetry and distribution of spin among the nucleon, magnetic moment of proton are some of the low energy properties for which the most of the researches are associated with. Along with proton, other members of baryon octet are also following the same symmetry, therefore, must follow the same criteria. Lambda, the lightest particle with strange quark is of motivation in this paper due to reconstruction of longitudinal and transverse polarization from non-leptonic decays. Parity violations in $\Lambda \rightarrow p\pi$ decays conserve polarization which makes it rare from other particles in baryon octet. Although, the SU(6) model helps to explain the spin distribution of the lambda but the predictions from the quark model are contradictory from the deep-inelastic experiments which claims the spin content from u and d quarks to be ≈40% and rest part may come from strange quark. Spin structure of Lambda is of special importance in enriching our knowledge of hadronic structure and hadronic spectrum. Due to polarization of u and d quarks and more delicacy towards SU(3) symmetry breaking help in quark fragmentation process which helps to go into deeper details of the hadronic puzzles.

The distribution of spin among its constituent quarks is studied through various models [1-4] which claims about the non-zero polarization of u, d and s quark. The correct values of spin transfer from the constituent quarks can help to enrich our knowledge of hadronic structure and spectrum through quark fragmentation process. Thus various phenomenologists have suggested different models [5-8] to study details of spin and quark content for Λ.

In section I, we use statistical modeling techniques to find the probability of all Fock state in lambda hyperon without any parameter. The principle of detailed balance is here applied to this lightest strange baryon with spin ½ from the pure statistical effects which were earlier introduced by Zhang et al. [9-11] to find the flavor asymmetry between u and d quark for proton. Section II uses the above mentioned principle to calculate the quark spin content of the hyperon, semileptonic decay constants and magnetic moment. Treating the hadrons as an ensemble of quark-gluon Fock states, the multiplicities calculated in spin and color spaces is further utilized to solve some of the puzzles regarding spin of lambda which is included in section III.

**Principle of detailed balance:**

The principle of detailed balance assumes that the composition of hadrons can be expanded by Fock states description in terms of the complete set of quark-gluon Fock states as

$$|h\rangle = \sum_{i,j,k} c_{i,j,l,k} |\{q\},\{i,j,l\},\{k\}\rangle$$

Where $\{q\}$ represents the valence quarks of the baryon, i is the number of quark- anti quark $u\bar{u}$ pairs, j is the number of quark-antiquark $d\bar{d}$ pairs, k is the number of gluons, and l is the number of $s\bar{s}$ pairs. Taking the hadronic system as an ensemble of Fock states where $\rho_{i,j,l,k}$ is the probability associated to find the quark-gluon Fock states

$$\rho_{i,j,l,k} = |c_{i,j,l,k}|^2$$

Where $\rho_{i,j,l,k}$ satisfies the normalization condition $\sum_{i,l,l,k} \rho_{i,j,l,k} = 1$.

The quarks and gluons in the Fock states are the intrinsic partons of the baryon which are multi-connected to the valence quarks. We here propose to apply principle of detailed balance for Lambda baryon to find the probability distributions and further used to solve various low energy parameters of lambda. Principle of detailed balance assumes the balancing of any two sub ensembles with each other and it can be written as:

$$\rho_{i,j,l,k}|\{q\},\{i,j,l.k\}\rangle \xrightleftharpoons{balance} \rho_{i',j',l',k'}|\{q\},\{i',j',l',k'\}\rangle$$

The same principle has been used by Zhang et al. [9] to explain quark flavor anti-symmetry of proton. Here we extend this principle for lambda and include the heavier flavor like $s\bar{s}$ pairs where as the original probability for proton do not include this flavor. The principle of detailed balance includes various sub processes like $g \Leftrightarrow q\bar{q}$, $g \Leftrightarrow gg$ and $q \Leftrightarrow qg$ are considered. The modification here comes with addition of $s\bar{s}$ pairs which are mainly generated from $g \Leftrightarrow s\bar{s}$ pairs.

(i) When $q \rightarrow qg$ is considered: Following the same procedure $s \rightarrow sg$ modifies the general expression of probability as:

$$|uds,\{i,j,l,k-1\}\rangle \underset{(3+2i+2j+2l)k}{\overset{3+2i+2j+2l}{\Leftrightarrow}} |uds\{i,j,l,k\}\rangle$$

Where i refer to $u\bar{u}$ pairs, j refer to $d\bar{d}$, l refers to $s\bar{s}$ and k refers to no. of gluons so that total number of partons are 3+2i+2j+2k+2l in the final state.

$$\frac{\rho_{i,j,l,k}}{\rho_{i,j,l,k-1}} = \frac{1}{k}$$

(ii) When both the processes $g \Leftrightarrow gg$ and $q \Leftrightarrow qg$ are included: Similarly,

$$|uds,\{i,j,l,k-1\}\rangle \xrightleftharpoons[(3+2i+2j+2l)k+\frac{k(k-1)}{2}]{3+2i+2j+2l+k-1} |uds,i,j,l,k\rangle$$

$$\frac{\rho_{i,j,l,k}}{\rho_{i,j,l,k-1}} = \frac{(3+2i+2j+2l+k-1)}{(3+2i+2j+2l)k+\frac{k(k-1)}{2}}$$

(iii) When $g \Leftrightarrow \bar{q}q$ is considered:- As the mass of strange quark is large, for gluons to undergo the process $g \Leftrightarrow s\bar{s}$, it must have free energy greater than at least two times the mass of strange quark, $\varepsilon_g > 2M_s$ where $M_s$ is the mass of strange quark. The generation of $s\bar{s}$ pair from gluons is restricted by applying a constraint in the form of $k(1-C_l)^{n-1}$ [12] where $n$ is the total number of partons present in the Fock state. The factor here is introduced from the gluon free energy distribution and applying some constraints to the momenta and total energy of partons present in the baryon. In all cases $C_{l-1} = \frac{2M_s}{M_\Lambda - 2(l-1)M_s}$, Ms is the mass of s-quark and $M_\Lambda$ is the mass of Lambda baryon.

$$|udsg\rangle \xrightleftharpoons[2X1]{1X(1-C_0)^3} |uds s\bar{s}\rangle$$

$$|uds\bar{s}sg\rangle \xrightleftharpoons[2X3]{1(1-C_1)^5} |uds\bar{s}ss\bar{s}s\rangle$$

$$|uds\bar{d}dg\rangle \xrightleftharpoons[1X2]{1(1-C_0)^5} |uds\bar{d}d s\bar{s}\rangle$$

$$|uds\bar{u}ug\rangle \xrightleftharpoons[1X2]{1(1-C_0)^5} |uds\bar{u}u s\bar{s}\rangle$$

$$|uds,i-1,j,l,k\rangle \xrightleftharpoons[i(i+1)]{k} |uds,i,j,l,k-1\rangle$$

Here for come in probability, i number of $\bar{u}$ pairs can combine with i+1 number of u quarks to give a single gluon. Similar expression can be written for j number of $\bar{d}d$ pairs.

$$|uds,i-1,j,l,1\rangle \xrightleftharpoons[i(i+1)]{1} |uds,i,j,l,1-1\rangle$$

$$\Rightarrow \frac{\rho_{i,j,l,0}}{\rho_{i-1,j,l,1}} = \frac{1}{i(i+1)}$$

$$|uds,i,j-1,l,1\rangle \xrightleftharpoons[j(j+1)]{1} |uds,i,j,l,1-1\rangle$$

$$\Rightarrow \frac{\rho_{i,j,l,0}}{\rho_{i,j-1,l,1}} = \frac{1}{j(j+1)}$$

$$\{|uds,i,j,l-1,k\rangle\} \xrightleftharpoons[l(l+1)]{k(1-C_{l-1})^{n-1}} \{|uds,i,j,l,k\rangle\}$$

$$\frac{\rho_{i,j,l,k}}{\rho_{i,j,l-1,k}} = \frac{k(1-C_{l-1})^{n-1}}{l(l+1)}$$

Here by replacing "l-1" by zero

$$|uds,i,j,0,k\rangle \xrightarrow{\frac{k(1-C_0)^{n-2l-1}}{1(1+1)}} |uds,i,j,1,k-1\rangle \text{ where } n = 3+2i+2j+2l+k$$

$$\frac{\rho_{i,j,1,k-1}}{\rho_{i,j,0,k}} = \frac{k(1-C_0)^{n-2l-1}}{1(1+1)}$$

$$|uds,i,j,1,k-1\rangle \xrightarrow{\frac{(k-1)(1-C_1)^{n-2l}}{2(2+1)}} |uds,i,j,2,k-2\rangle$$

$$\frac{\rho_{i,j,2,k-2}}{\rho_{i,j,1,k-1}} = \frac{(k-1)(1-C_1)^{n-2l}}{2(2+1)}$$

$$|uds,i,j,2,k-2\rangle \xrightarrow{\frac{(k-2)(1-C_2)^{n-2l+1}}{3(3+1)}} |uds,i,j,3,k-3\rangle$$

$$\frac{\rho_{i,j,3,k-3}}{\rho_{i,j,2,k-2}} = \frac{(k-2)(1-C_2)^{n-2l+1}}{3(3+1)}$$

Here the go-out probability depends upon the number of partons present at that time. Similarly if we proceed until

$$|uds,i,j,k-1,1\rangle \xrightarrow{\frac{1(1-C_{k-1})^{n-k-2}}{k(k+1)}} |uds,i,j,k,o\rangle$$

$$|\rho_{i,j,k,0}\rangle \xrightarrow{\frac{1(1-C_{k-1})^{n-k-2}}{k(k+1)}} |\rho_{i,j,k-1,1}\rangle$$

Thus

$$\frac{\rho_{i,j,l,0}}{\rho_{i,j,0,l}} = \frac{(k(k-1)(k-2)(k-3)---1(1-C_0)^{n-2l-1}(1-C_1)^{n-2l}(1-C_2)^{n-2l+1}(1-C_{l-1})^{n-k-2})}{k!(k+1)!}$$

$$|uds,i,j,l+k-1,1\rangle \xrightarrow{\frac{1(1-C_{l-1})^{n-k-2}}{l+k(l+k+1)}} |uds,i,j,l+k,0\rangle \text{ where } n-k-2 = 3+2i+2j+2l-2.$$

$$\frac{\rho_{i,j,l,k}}{\rho_{i,j,l+k,0}} = \frac{(k(k-1)(k-2)(k-3)---1(1-C_0)^{n-2l-1}(1-C_1)^{n-2l}(1-C_2)^{n-2l+1}(1-C_{l-1})^{n-k-2})}{(l+1)(l+2).......(l+k)(l+k+1)}$$

$$\frac{\rho_{i,j,l+k,0}}{\rho_{0,0,0,0}} = \frac{1}{i!\,i+1!\,j!(j+1)!(l+k)!(l+k+1)!}$$

Thus, all the probabilities can be written in terms of $\rho_{0,0,0,0}$ which can be solved by applying the constraint by taking sum of all probabilities as one.

**The entire list of $\rho_{i,j,l,k}$'s can be obtained and shown in table 1.1.**

| i | j | l | $\rho_{i,j,l}$ | k=0 | k=1 | k=2 |
|---|---|---|---|---|---|---|
| 0 | 0 | 0 | $uds$ | 0.12335 | 0.0319116 | 0.051238 |
| 0 | 0 | 1 | $uds s\bar{s}$ | 0.061675 | 0.061675 | 0.0336409 |
| 0 | 1 | 0 | $uds d\bar{d}$ | 0.061675 | 0.0102836 | 0.0165116 |
| 1 | 0 | 0 | $uds u\bar{u}$ | 0.061675 | 0.0102836 | 0.0165116 |
| 0 | 1 | 1 | $uds d\bar{d} s\bar{s}$ | 0.0308375 | 0.0308375 | 0.0164467 |
| 1 | 0 | 1 | $uds u\bar{u} s\bar{s}$ | 0.0308375 | 0.0308375 | 0.0164467 |
| 1 | 1 | 0 | $uds u\bar{u} d\bar{d}$ | 0.0308375 | 0.00331392 | 0.0053209 |
| 1 | 1 | 1 | $uds u\bar{u} d\bar{d} s\bar{s}$ | 0.0154188 | 0.0154188 | 0.00811513 |
| 1 | 2 | 0 | $uds u\bar{u} d\bar{d} d\bar{d}$ | 0.00513958 | 0.000355973 | 0.000571558 |
| 0 | 2 | 0 | $uds d\bar{d} d\bar{d}$ | 0.0102792 | 0.00110464 | 0.00177363 |
| 0 | 2 | 1 | $uds d\bar{d} d\bar{d} s\bar{s}$ | 0.00513958 | 0.00513958 | 0.00270504 |
| 2 | 0 | 0 | $uds u\bar{u} u\bar{u}$ | 0.0102792 | 0.00110464 | 0.00177363 |
| 2 | 1 | 0 | $uds u\bar{u} u\bar{u} d\bar{d}$ | 0.00513958 | 0.000355973 | 0.000571558 |
| 2 | 0 | 1 | $uds u\bar{u} u\bar{u} s\bar{s}$ | 0.00513958 | 0.00513958 | 0.00270504 |
| 0 | 3 | 0 | $uds d\bar{d} d\bar{d} d\bar{d}$ | 0.000856597 | 0.0000593288 | 0.000095296 |
| 3 | 0 | 0 | $uds u\bar{u} u\bar{u} u\bar{u}$ | 0.000856597 | 0.0000593288 | 0.000045296 |

Here we have assumed that there exists equal probability for each Fock state to reside inside the hadron but partons are not free particles. The momenta associated with each Fock state puts some conditions on the parton distribution of the hadron. Thus the number of strange quark antiquark pairs have been limited to atmost one for $\rho_{i,j,l,k}$'s due its large mass and limited free energy of a gluon undergoing the sub process $g \Leftrightarrow s\bar{s}$. Therefore constrain on $s\bar{s}$ pairs results in symmetry in probability values for u and d quarks only.

**Statistical Model with its application to Lambda Hyperon:**

Assuming hadron as an ensemble of quark-gluon Fock states based on the same principle, various properties of lambda can be found out. The quarks and gluons are multi-connected to valence quarks and must conserve the intrinsic properties of the hadron itself. The wave-function of a hadronic system which includes three quarks known as valence part and sea-content can be written in terms of spin, flavor and color of valence and sea separately. Thus various combinations for $q^3$ and sea wave-function so as to give a color singlet and spin half, flavor octet state are given below. Each combination maintains the antisymmetry of the complete hadronic wave-function.

$\Phi_1^{\frac{1}{2}} H_0 G_1$, $\Phi_8^{\frac{1}{2}} H_0 G_8$, $\Phi_{10}^{\frac{1}{2}} H_0 G_{\overline{10}}$, $\Phi_1^{\frac{1}{2}} H_1 G_1$, $\Phi_8^{\frac{1}{2}} H_1 G_8$, $\Phi_{10}^{\frac{1}{2}} H_1 G_{\overline{10}}$ and $\Phi_8^{\frac{3}{2}} H_1 G_8$, $\Phi_8^{\frac{3}{2}} H_2 G_8$

The wave-function for the valence part is denoted by $\Psi = \Phi[|\phi\rangle|\chi\rangle|\psi\rangle]|\xi\rangle$ where $|\Phi\rangle$ denotes the flavor part of $q^3$ wave-function for baryons, $|\chi\rangle$ represents spin of valence part in the wave-function, $|\Psi\rangle$ is for color and $|\xi\rangle$ is for space contribution to the whole wave-function and $H_{0,1,2}$ and $G_{1,8,10}$ denote the sea-part. The total flavor-spin-color wave function of a spin up baryon which consists of three-valence quarks and sea components can be written as given in [13-14].

$$|\Phi_{\frac{1}{2}}^{\uparrow}\rangle = \frac{1}{N}[\Phi_{1}^{(\frac{1}{2})^{\uparrow}} H_0 G_1 + a_8 \Phi_{8}^{(\frac{1}{2})^{\uparrow}} H_0 G_8 + a_{10} \Phi_{10}^{(\frac{1}{2})^{\uparrow}} H_0 G_{\overline{10}} + b_1 \left[\Phi_{1}^{\frac{1}{2}} \otimes H_1\right]^{\uparrow} G_1 +$$

$$+ b_8 \left(\Phi_{8}^{\frac{1}{2}} \otimes H_1\right)^{\uparrow} G_8 + b_{10} \left(\Phi_{10}^{\frac{1}{2}} \otimes H_1\right)^{\uparrow} G_{\overline{10}} + c_8 (\Phi_{8}^{\frac{3}{2}} \otimes H_1)^{\uparrow} G_8 + d_8 (\Phi_{8}^{\frac{3}{2}} \otimes H_2)^{\uparrow} G_8](1.6)$$

Where $\quad N^2 = 1 + a_8^2 + a_{10}^2 + b_1^2 + b_8^2 + b_{10}^2 + c_8^2 + d_8^2$

Wave-functions $\Phi_{b1}^{(\frac{1}{2})^{\uparrow}}, \Phi_{b8}^{(\frac{1}{2})^{\uparrow}}, \Phi_{b10}^{(\frac{1}{2})^{\uparrow}}, \Phi_{c8}^{(\frac{1}{2})^{\uparrow}}, \Phi_{d8}^{(\frac{1}{2})^{\uparrow}}$ are written by taking coupling between spins of sea part and flavor-spin part of $q^3$ wave-function along with suitable normalization constant. The sea and core part has been considered to be in S-wave and only color and spin of the sea-part is mainly focused and assumed to be flavorless. However another possibility is to assume sea to be colorless and having flavor and spin.[15-16].

Now the statistical decomposition of hadrons in quark-gluon Fock states $|u, d, s, i, j, l, k\rangle$ can be taken in terms of sub-processes $q \Leftrightarrow qg$, $g \Leftrightarrow gg$ and $g \Leftrightarrow q\bar{q}$. The statistical model which has been applied on nucleonic system earlier by Singh and Upadhyay [17] and it involves computation of probability $\rho_{j_1, j_2}$ where relative probability for the core quarks to have spin $j_1$ and the same for gluons to have spin $j_2$ and then probability for $j_1$ and $j_2$ so as to give resultant probability as ½. Thus the product probability ratios are expressed in the form of a common parameter c. "c" has its own significance in the sense that parameters $\alpha$ and $\beta$ are computed using the value of c where c is multiplied with a suitable multiplicity factor n. Moreover sea is assumed to be in S-wave and thus relativistic effects has been neglected first.

**The table for "nc" is shown below:**

| States | $H_0 G_1$ | $H_0 G_8$ | $H_0 G_{\overline{10}}$ | $H_1 G_1$ | $H_1 G_8^{\frac{1}{2}}$ | $H_1 G_{\overline{10}}$ | $H_2 G_8$ | $H_1 G_8^{\frac{3}{2}}$ |
|---|---|---|---|---|---|---|---|---|
| $|gg\rangle$ | 0.0064075 | 0.0128095 | 0 | 0 | 0.0128095 | 0.00640475 | 0.00640475 | 0.00640475 |
| $|\bar{u}ug\rangle$ | 0.000642725 | 0.0025709 | 0.000642725 | 0.000642725 | 0.0025709 | 0.000642725 | 0.00128545 | 0.00128545 |

| | | | | | | | | |
|---|---|---|---|---|---|---|---|---|
| $\|\bar{d}dg\rangle$ | 0.000642725 | 0.0025709 | 0.000642725 | 0.000642725 | 0.0025709 | 0.000642725 | 0.00128545 | 0.00128545 |
| $\|\bar{s}sg\rangle$ | 0.0038549 | 0.0154188 | 0.0038549 | 0.0038549 | 0.0154188 | 0.0038549 | 0.00770938 | 0.00770938 |
| $\|\bar{u}u\bar{d}d\rangle$ | 0.00192734 | 0.00770938 | 0.00192734 | 0.00192734 | 0.0770938 | 0.00192734 | 0.00385469 | 0.00385469 |
| $\|\bar{u}u\bar{s}s\rangle$ | 0.00192734 | 0.00770938 | 0.00192734 | 0.00192734 | 0.00770938 | 0.00192734 | 0.00385469 | 0.00385469 |
| $\|\bar{d}d\bar{s}s\rangle$ | 0.00192734 | 0.00770938 | 0.00192734 | 0.00192734 | 0.00770938 | 0.00192734 | 0.00385469 | 0.00385469 |
| $\|\bar{d}d\bar{d}d\rangle$ | 0.0012849 | 0.0025698 | 0 | 0 | 0.0025698 | 0.0012849 | 0.0012849 | 0.0012849 |
| $\|ggg\rangle$ | 0.00480791 | 0.00961583 | 0 | 0.0080791 | 0.0096158 | 0 | 0 | 0 |
| $\|\bar{u}u\bar{d}dg\rangle$ | 0.00005178 | 0.00041424 | 0.0103756 | 0.00015534 | 0.00124272 | 0.00031068 | 0.00041424 | 0.00062136 |
| $\|\bar{u}u\bar{s}sg\rangle$ | 0.00047476 | 0.00379797 | 0.000949492 | 0.00142424 | 0.0113939 | 0.00284848 | 0.00379797 | 0.00569695 |
| $\|\bar{d}d\bar{s}sg\rangle$ | 0.000481836 | 0.00385469 | 0.000963672 | 0.00014451 | 0.0115640625 | 0.00289102 | 0.00385469 | 0.00578203 |
| $\|\bar{d}d\bar{d}dg\rangle$ | 0.000184107 | 0.000147285 | 0.0000368213 | 0.000055232 | 0.0000441856 | 0.000110464 | 0.0000736427 | 0.00022092 |
| $\|\bar{u}ugg\rangle$ | 0.000275193 | 0.00220155 | 0.000550387 | 0.00082558 | 0.00660464 | 0.00165116 | 0.00110077 | 0.00330232 |
| $\|\bar{d}dgg\rangle$ | 0.00275193 | 0.00220155 | 0.000550387 | 0.00082558 | 0.00660464 | 0.00165116 | 0.00110077 | 0.00330232 |
| $\|\bar{s}sgg\rangle$ | 0.000560673 | 0.00448539 | 0.00112135 | 0.0168202 | 0.0134562 | 0.00336404 | 0.00224269 | 0.00672808 |
| $\|\bar{u}u\bar{u}u\rangle$ | 0.0012849 | 0.0025698 | 0 | 0 | 0.0025698 | 0.0012849 | 0.0012849 | 0.0012849 |
| $\|\bar{u}u\bar{u}ug\rangle$ | 0.0000184107 | 0.000147285 | 0.0000368213 | 0.000055232 | 0.000491856 | 0.000110464 | 0.0000736427 | 0.000220928 |
| $\|0\rangle$ | 0.12335 | - | -- | -- | -- | -- | -- | -- |
| $\|g\rangle + \|\bar{u}u\rangle + \|\bar{d}d\rangle + \|\bar{s}s\rangle$ | | | | | 0.1446264 | | | 0.0723132 |
| Total | 0.152011 | 0.0885036 | 0.0152347 | 0.00221988 | 0.267628 | 0.0328342 | 0.043413 | 0.129006 |

The total flavor-spin-color wave function of a spin up baryon when summing over all the probabilities over all the states given above produce five parameters, a = 0.475369, b = -0.0017437, c=0.0142791, d=0.190324, e = 0.583127 used to calculate physical quantities related to the spin of a nucleon and computes two parameters, $\alpha$ =0.2089, $\beta$ =0.067431.

**Discussion of the spin distribution and related properties:**

Naïve quark model predicts that all the spin is carried by s-quark, the contribution from u, d quarks are zero but the predictions from this model failed to explain the DIS experiments.
The total contribution of spin among hadrons is given by integration of g1 over x and called first moment which can be written as:

$$\Gamma_1 = \int_0^1 g1(x)dx = \frac{1}{2}\int_0^1 \sum_f e_f^2 \Delta q_f(x)dx$$

In application, we need to find the quantities, $\Delta q = n(q\uparrow) - n(q\downarrow) + n(\bar{q}\uparrow) - n(\bar{q}\downarrow)$ for q=u, d, s where $n(q\uparrow)$ is the number of spin up and $n(q\downarrow)$ is the number of spin down quarks of flavor q in spin-up baryon and same meaning can be understood for anti quarks. Integrate polarized quark densities can be used to define coupling constants for a baryon as:

$$g_A^{(0)} = \Delta U + \Delta D + \Delta S$$

$$g_A^{(3)} = \Delta U - \Delta D$$

$$\sqrt{3} g_A^8 = \Delta U + \Delta D - 2\Delta S$$

In SU(3) limit, $\Delta U = \Delta D$ gives $g_A^{(3)} = 0$. Applying the QCD corrections the above expression can be modified by addition of scale dependence ($Q^2$) and modified expression becomes $\Gamma_1(Q^2) = \int_0^1 g1(x, Q^2) dx = \frac{1}{18}(4\Delta U + \Delta D + \Delta S)(1 - \alpha_s(\pi))$ where $(1 - \alpha_s \pi)$ is the first order corrections derived from Bjorken sum rule [18]. F and D are the SU(3) invariant matrix elements obtained from the hyperon beta decay and one needs to express all the parameters in terms of these matrix elements. Also total spin content is given by $\Delta \Sigma = \Delta U + \Delta D + \Delta S = g_A^{(0)}$. In terms of $\Delta \Sigma$, it can be stated that $\Delta U = \Delta D = \frac{1}{3}(\Delta \Sigma - D)$, $\Delta S = \frac{1}{3}(\Delta \Sigma + 2D)$ and $\Gamma_1^\Lambda = \frac{1}{18}(2\Delta \Sigma - D)$.

In order to include sea contribution, we need to write the above defined parameters in our model as: $\Delta U = \Delta D = \frac{\alpha}{2} - 2\beta = -0.03$ and $\Delta S = 2\alpha + \beta = 0.49$. Assuming SU(3) invariance, $F = 3\alpha/2$ and $D = 3(\alpha + 2\beta)/2$. We use F and D again to find the vector and axial coupling constant ratio for semi leptonic decay of and $g_A^{(8)} = -2D/\sqrt{3}$. For spin distribution of lambda, $\Gamma_1^\Lambda = \frac{1}{6}(\alpha - 4\beta)$ is the required expression. A table shown below lists calculated value for all the above mentioned properties and some more results from other phenomenological models has also been listed here.

**Table showing the comparison of calculated low energy parameters with other models:**

| Lambda | Statistical Model | Data | Ref.[20] | Ref.[21] |
|---|---|---|---|---|
| $\Delta U$ | -0.03 | -0.02±0.04 [25] | -0.03±0.14 | - |
| $\Delta D$ | -0.03 | -0.02±0.04 [25] | -0.03±0.14 | - |
| $\Delta S$ | 0.49 | 0.68±0.04 [25] | 0.74±0.17 | - |
| $\Gamma_1^\Lambda$ | 0.019 | - | - | 0.027 |
| $g_A^{(8)}$ | -0.595 | 0.718±0.015[19] | - | - |
| $F/D$ | 0.607 | 0.575[19] | - | 0.733 |
| $\Sigma = \Delta U + \Delta D + \Delta S$ | 0.43 | - | 0.68±0.44 | - |

**Discussion of Result:**

Our model here calculates all the low energy parameters in a statistical frame. The principle used here is successful in explaining the flavor asymmetry of nucleon [9, 10] and application to this principle in the form of statistical model [17] also produced authentic results. Now we have extended this to apply the model to lambda where the contributions from valence as well as sea part are taken from two parameters α and β as stated earlier. The table 1.1 shown above can be justified from the less probability outcome for higher mass Fock states and follows u and d isospin symmetry. The calculations here are performed in baryonic rest frame. Our result for spin distribution of lambda matches with value $\int dx g_1^\Lambda(x) = 0.022 \pm 0.014$ [22] which is extracted by solving for current $\langle N | \delta \overline{\gamma_\mu} \gamma_5 | N \rangle$. Although SU(6) predicts spin of lambda to be 1/18 but our model predicts values less than its SU(6). From EMC data of spin distribution of proton, $\Gamma_1^\Lambda = 0.42 \pm 0.19$ can be estimated [24]. All parameters in ref.[20] modifies to $\Delta U = \Delta D = -0.02 \pm 0.17$ and $\Delta S = 1.21 \pm 0.54$ [23] when strange quark mass corrections are applied. A comparison with experimental data and other models show that results produced from statistical model are close to experimental data [19] and other models. The assumption of non-relativistic nature of quarks and sea inside a baryon affects the results very little as compared to Ref. [21] where the parameters are rescaled to include orbital motion of quarks. The unavailability of -targets leads to insufficient experimental data on lambda hyperon. The statistical model used here covers 75% of the Fock states. The suitable reason for this comes from constraint applied on number of $s\overline{s}$ pairs.